# Interface ferromagnetism and anomalous Hall effect of CdO/ferromagnetic insulator heterostructures

Yang Ma[1,2,†], Yu Yun[1,2,†], Yuehui Li[1,3], Wenyu Xing[1,2], Yunyan Yao[1,2], Ranran Cai[1,2], Yangyang Chen[1,2], Yuan Ji[1,2], Peng Gao[1,2,3], Xin-Cheng Xie[1,2,4,5], and Wei Han[1,2,*]

[1]International Center for Quantum Materials, School of Physics, Peking University, Beijing 100871, People's Republic of China

[2]Collaborative Innovation Center of Quantum Matter, Beijing 100871, People's Republic of China

[3]Electron Microscopy Laboratory, School of Physics, Peking University, Beijing, 100871, People's Republic of China

[4]CAS Center for Excellence in Topological Quantum Computation, University of Chinese Academy of Sciences, Beijing 100190, People's Republic of China

[5]Beijing Academy of Quantum Information Sciences, Beijing 100193, People's Republic of China

[†]These authors contributed equally to the work
[*]Correspondence to: weihan@pku.edu.cn

The experimental observation of quantum anomalous Hall effect (QAHE) in magnetic topological insulators has stimulated enormous interest in condensed-matter physics and materials science. For the purpose of realizing high-temperature QAHE, several material candidates have been proposed, among which the interface states in the CdO/ferromagnetic insulator heterostructures are particularly interesting and favorable for technological applications. Here, we report the experimental observation of the interfacial ferromagnetism and anomalous Hall effect in the $Fe_3O_4$/CdO/$Fe_3O_4$ heterostructures grown via oxide molecular-beam epitaxy. Systematical variation of the CdO thickness reveals the interface

**ferromagnetism as the major cause for the observed planar magnetoresistance and anomalous Hall effect. Our results might pave the way to engineer oxide interface states for the exploration of QAHE towards exotic quantum-physical phenomena and potential applications.**

## I. INTRODUCTION

The adding of quantum anomalous Hall effect (QAHE) to the Hall family has stimulated enormous interest in the field of condensed-matter physics and material science [1-3]. The experimental observation of QAHE without any magnetic field has been achieved in a magnetic topological insulator (TI) Cr-doped and V-doped $(Bi,Sb)_2Te_3$ thin films grown by molecular-beam epitaxy at low temperature [4-9]. The QAHE states hold a variety of exotic quantum phenomena, including topological magnetoelectric effect [10-12], magnetic monopoles [13], chiral Majorana fermion modes [14], etc. To fulfill these potentials towards technological applications, one of the major challenges is to increase the critical temperature of QAHE with quantized Hall resistance, which is still ~1 K to date [7,15,16]. For the purpose of robust and high-temperature QAHE, a large variety of material candidates has been proposed and/or experimentally studied, including *n-p* co-doped magnetic TIs [17], magnetic proximitized TIs [18-21], graphene/ferromagnetic insulator heterostructures [22,23], oxide interfaces [24,25], as well as magnetic silicene and single-layer tin [26-28], etc. Among all these candidates, the oxide interface states in CdO/ferromagnetic insulator (FI) heterostructures are particularly interesting since the interface is protected by two oxide layers leading to the robustness of QAHE [24]. Both the topological properties and the proximity-effect-induced ferromagnetism give rise to the expected QAHE. Thus, the experimental exploration of the interface ferromagnetism and the electron-transport properties in CdO/FI heterostructures is emergent.

In this paper, we report the experimental observation of interface ferromagnetism and anomalous Hall effect (AHE) in the FI/CdO/FI trilayer heterostructures grown by oxide molecular-beam epitaxy. The FI layer is made of thin $Fe_3O_4$ film instead of EuO as proposed in previous theoretical work [24], due to the extremely different oxygen partial pressures needed for the growth of CdO and EuO thin films [29,30]. The interface ferromagnetism in the $Fe_3O_4$/CdO/$Fe_3O_4$ trilayer heterostructures is characterized via both the AHE and planar magnetoresistance (MR) measurements. Systematic studies of the CdO thickness dependence of AHE and planar MR reveal the most important role of interface ferromagnetism in the $Fe_3O_4$/CdO/$Fe_3O_4$ heterostructures. Furthermore, the major cause for the observed anomalous Hall resistivity far from the quantized resistivity has been identified, namely the oxygen vacancies in the CdO layer, which contribute to the bulk conducting carriers. The observed interfacial ferromagnetism fulfills one of the prerequisites for realizing QAHE, while the topological properties need further studies. Our results could be important for engineering oxide interface states for the exploration of QAHE.

## II. EXPERIMENTAL

Figure 1(a) illustrates the $Fe_3O_4$/CdO/$Fe_3O_4$ trilayer heterostructures, where the $Fe_3O_4$ is the ferromagnetic insulating layer with a fixed thickness of 4 unit cells (UC), and the CdO is the band-insulating layer with thickness varied to tune the electron-transport and magnetic properties. The $Fe_3O_4$/CdO/$Fe_3O_4$ trilayer heterostructures are grown on (001)-oriented MgO substrates via an oxide molecular-beam expitaxy system (MBE-Komponenten GmbH; Octoplus 400) with a base pressure lower than $5\times10^{-10}$ mbar. Prior to the growth, the MgO substrates are pre-cleaned by annealing at 600 ℃ for 2 h. The $Fe_3O_4$ is grown by evaporating Fe from a thermal effusion cell (rate: 0.02 Å/s) in diluted ozone gas ($O_3$:~11%) under a pressure of $5.2\times10^{-7}$ mbar. The CdO layer is grown by evaporating Cd from a thermal effusion cell (rate:0.03 Å/s) in diluted ozone gas ($O_3$:~18%) under a pressure of $4.3\times10^{-5}$ mbar. During the growth of the heterostructures, *in situ* integrated reflective high-energy electron diffraction (RHEED) is used to monitor the crystalline properties of each layer. Figs. 1(b)-1(e) show the RHEED patterns of the MgO substrate, the bottom $Fe_3O_4$ layer (4 UC), the CdO layer (15 UC), and the top $Fe_3O_4$ layer (4 UC), respectively, viewed from the MgO crystal's [100] direction. Based on the line cuts of the RHEED images, the strain strengths are estimated to be 2.33% (compressive), 3.98% (compressive), and 2.33% (compressive) for the bottom $Fe_3O_4$, CdO, and top $Fe_3O_4$, respectively (Supplemental Material, Fig. S1 and Table S1) [31]. Prior to moving the $Fe_3O_4$/CdO/$Fe_3O_4$ trilayer samples out of the high-vacuum chamber, an ~3-nm MgO thin film is deposited via *e*-beam evaporation to avoid degradation during the characterization.

The interface quality of the $Fe_3O_4$/CdO/$Fe_3O_4$ trilayer samples is further characterized by scanning transmission electron microscopy (STEM). The high-angle annular dark-field images (Figs. 1(f) and 1(g)) are recorded at 300 kV using an aberration-corrected FEI Titan Themis G2 with spatial resolutions up to ~65 pm. The sharp interface of the atomically resolved *Z*-contrast (*Z* is atomic number) image does not exhibit any signature of intermixing of the Cd and Fe atoms across the interface (Fig. 1(f)). The sharp interface between CdO and $Fe_3O_4$ layers is further confirmed by the energy-dispersive x-ray spectroscopy (EDS) via using Bruker Super-X detectors, as shown in Figs. 1(h) and 1(i). Both RHEED and STEM characterizations show the good crystalline quality of the $Fe_3O_4$/CdO/$Fe_3O_4$ trilayer heterostructures.

The magnetization measurement of the $Fe_3O_4$/CdO/$Fe_3O_4$ trilayer samples is performed in a Magnetic Properties Measurement System (MPMS; Quantum Design). The magnetization easy axis is in-plane (Supplemental Material, Fig. S2) [31]. The electron-transport measurement is performed in a Physical Properties Measurement System (PPMS; Quantum Design).

## III. RESULTS and DISCUSSION

To probe the interface ferromagnetism of the FI/CdO/FI trilayer heterostructures predicted by Zhang, *et al.* [24], we first perform the AHE measurements on the $Fe_3O_4$/CdO/$Fe_3O_4$ heterostructures consisting of 15-UC CdO film in a PPMS. The heterostructures are first fabricated to be Hall bar geometry (channel width: 100 $\mu$m; channel length: 1500 $\mu$m) via standard photolithography and etching process in diluted hydrochloric acid solution. During the transport measurements, a DC current of ~100 $\mu$A was applied via Keithley K2400 and the voltages were measured via Keithley K2002. Fig. 2(a) shows the sheet resistance as a function of the temperature for the Hall bar devices made on $Fe_3O_4$/CdO (15-UC)/$Fe_3O_4$ heterostructures. The much lower sheet resistivity of $Fe_3O_4$/CdO (15-UC)/$Fe_3O_4$ compared to 8-UC $Fe_3O_4$ film (Top panel) clearly shows that the transport properties measured on the heterostructures at low temperatures ($T <= 50$ K) are purely from the conducting CdO and the interface states. Fig. 2(b) depicts the schematic of magnetic proximity effect due to exchange interaction at the interface between CdO and $Fe_3O_4$. The overlap of wave functions between the Cd's 5*s* electrons and the spin-polarized Fe's 3*d* electrons at the localized $e_g$ band gives rise to the spin splitting of the Cd's 5*s* electrons in the conduction band. Fig. 2(c) shows the anomalous Hall resistance (bottom panel) as a function of the perpendicular magnetic field at *T*=10 K after subtracting the linear background arising from ordinary Hall effect. The same characteristic magnetic fields for AHE and magnetization (Top panel) of $Fe_3O_4$/CdO (15 UC)/$Fe_3O_4$ proves that the AHE arises from interface ferromagnetism due to the interfacial magnetic proximity effect between CdO and $Fe_3O_4$ [18,19,32]. The single-step magnetization switching indicates the ferromagnetic exchange coupling between two $Fe_3O_4$ layers. Since the CdO layer is nonmagnetic, where only linear ordinary Hall effect is observed [28], and the $Fe_3O_4$ layers are insulating, the observed AHE in $Fe_3O_4$/CdO/$Fe_3O_4$ structures can only be attributed to interface ferromagnetism as a result of magnetic proximity effect. The anomalous Hall resistivity exhibits little variation as the temperature changes between 2 and 50 K (Supplemental Material, Fig. S3) [31].

To further probe the interface ferromagnetism of the $Fe_3O_4$/CdO (15-UC)/$Fe_3O_4$ heterostructures, we measure the planar magnetoresistance (MR) on the Hall bar device with the magnetic field in the sample's plane. As shown in Fig. 3(a), the MR curves as a function of in-plane magnetic field that is perpendicular to the current direction (blue curves) clearly exhibit two MR maxima with a MR ratio ($\Delta\rho/\rho_{B=0}$) of $2.5\times10^{-4}$ at the in-plane coercive magnetic fields ($B_{peak}$). A hysteresis feature of the planar MR is observed, supporting the induced ferromagnetism at $Fe_3O_4$/CdO interfaces due to magnetic exchange interaction. As the in-plane magnetic field is swept parallel to the dc current, similar results (red curves) are observed. The comparable hysteretic MR maxima observed in these two measurement geometries could be associated with the extra scatterings for the conducting electrons in the presence of ferromagnetic domain walls around the in-plane coercive magnetic fields. These planar MR results are similar to previously reported magnetic TI induced by proximity effect with EuS, a ferromagnetic insulator [18]. The temperature dependence of the MR ratio (Fig. 3(b)) further rules out the origin of the observed AHE and MR due to any charges trapped in $Fe_3O_4$. If there are charges trapped in the $Fe_3O_4$ layer, the planar MR is expected to increase for increasing temperature since the conductance in $Fe_3O_4$ increases at elevated temperatures.

Next, we systematically vary the CdO layer thickness to study its effect on the interface ferromagnetism. Fig. 4(a) shows the temperature dependence of channel resistivity, and a semiconducting-behavior is observed for all these four samples (see Supplemental Material, Fig. S4 for the $Fe_3O_4$/CdO (15-UC)/$Fe_3O_4$ sample) [31]. The thickness dependence of the channel resistivity and the carrier density are shown in Supplemental Material, Fig. S5 [31]. As the CdO thickness increases from 6 to 15 UC, the MR ratio ($\Delta\rho/\rho_{B=0}$) of the $Fe_3O_4$/CdO/$Fe_3O_4$ heterostructures quickly decreases from ~$7.4\times10^{-3}$ to ~$2.2\times10^{-4}$ (Fig. 4(b) and Supplemental Material, Fig. S6) [31]. Fig. 4(c) shows anomalous Hall resistance curves as a function of perpendicular magnetic field measured on the $Fe_3O_4$/CdO (10-UC, 20-UC)/$Fe_3O_4$ heterostructures. A smaller anomalous Hall resistance is observed on $Fe_3O_4$/CdO (20-UC)/$Fe_3O_4$ heterostructures compared to $Fe_3O_4$/CdO (10-UC)/$Fe_3O_4$ heterostructures. Fig. 4(d) shows the CdO thickness dependence of the anomalous Hall conductivity to quantitatively investigate the role of interface ferromagnetism for the AHE. Since the magnetized interfaces and nonmagnetic bulk CdO channel are effectively in parallel connection, more current flows through nonmagnetic bulk channel for thicker CdO samples, which will reduce total AHE. The decrease of AHE conductivity upon increasing CdO thickness further supports that the observed AHE is due to interfacial ferromagnetism.

It is noted that the anomalous Hall resistivity measured in the $Fe_3O_4$/CdO/$Fe_3O_4$ heterostructures is several orders smaller compared to the QAHE with quantized resistivity [4]. We believe that the major cause is the current shunting effect due to bulk conducting states in the CdO layer associated with oxygen vacancies [29,33]. To minimize the population of oxygen vacancies, we have grown the samples by lowering the Cd evaporation rates in the same diluted ozone environment during the growth and post-annealing in oxygen environment after the growth. The postannealing is performed at the temperature of 300 °C to avoid the deformation of $Fe_3O_4$ to $Fe_2O_3$. Despite intensive effort, only a slight increase of the anomalous Hall resistivity is observed with a slight decrease of the carrier density (Supplemental Material, Fig. S7) [31]. This result could be associated with the easy formation of oxygen vacancies in CdO thin films, as reported in previous studies [29,33]. Nevertheless, the observation of AHE and planar MR in $Fe_3O_4$/CdO/$Fe_3O_4$ heterostructures identifies the interface ferromagnetism, which is the first major step to explore oxide interface for robust high-temperature QAHE. Using advanced growth techniques, i.e., oxide molecular-beam epitaxy with a pure ozone source ($O_3$:100%) [34], oxygen-vacancy-free CdO thin films could be achieved towards robust high-temperature QAHE.

## VI. CONCLUSION

In summary, we have grown the $Fe_3O_4$/CdO/$Fe_3O_4$ heterostructures via oxide molecular beam epitaxy and observed the interface ferromagnetism and anomalous Hall effect on the $Fe_3O_4$/CdO/$Fe_3O_4$ heterostructures. The observed interfacial ferromagnetism is a major step to explore oxide interface for robust high-temperature QAHE, while the topological properties need further studies. One major obstacle to identify the topological properties is related to the bulk conduction arising from the oxygen vacancies in CdO layer. Continuously improving the quality of CdO layer to achieve oxygen-vacancy-free CdO thin films could pave the way to experimentally probe the expected topological properties predicted by Zhang *et al.* [24]. Our experimental results could pave the way for engineering oxide interface states for the exploration of QAHE towards exotic quantum-physical phenomena and potential applications.

**ACKNOWLEDGMENTS**

We acknowledge the financial support from National Basic Research Programs of China (Grant No. 2015CB921104), National Natural Science Foundation of China (Grant No. 11574006, No. 11704011, No. 51502007 and 51672007), Beijing Natural Science Foundation (Grant No. 1192009), and the Key Research Program of the Chinese Academy of Sciences (Grant No. XDB28020100).

**Figure 1**

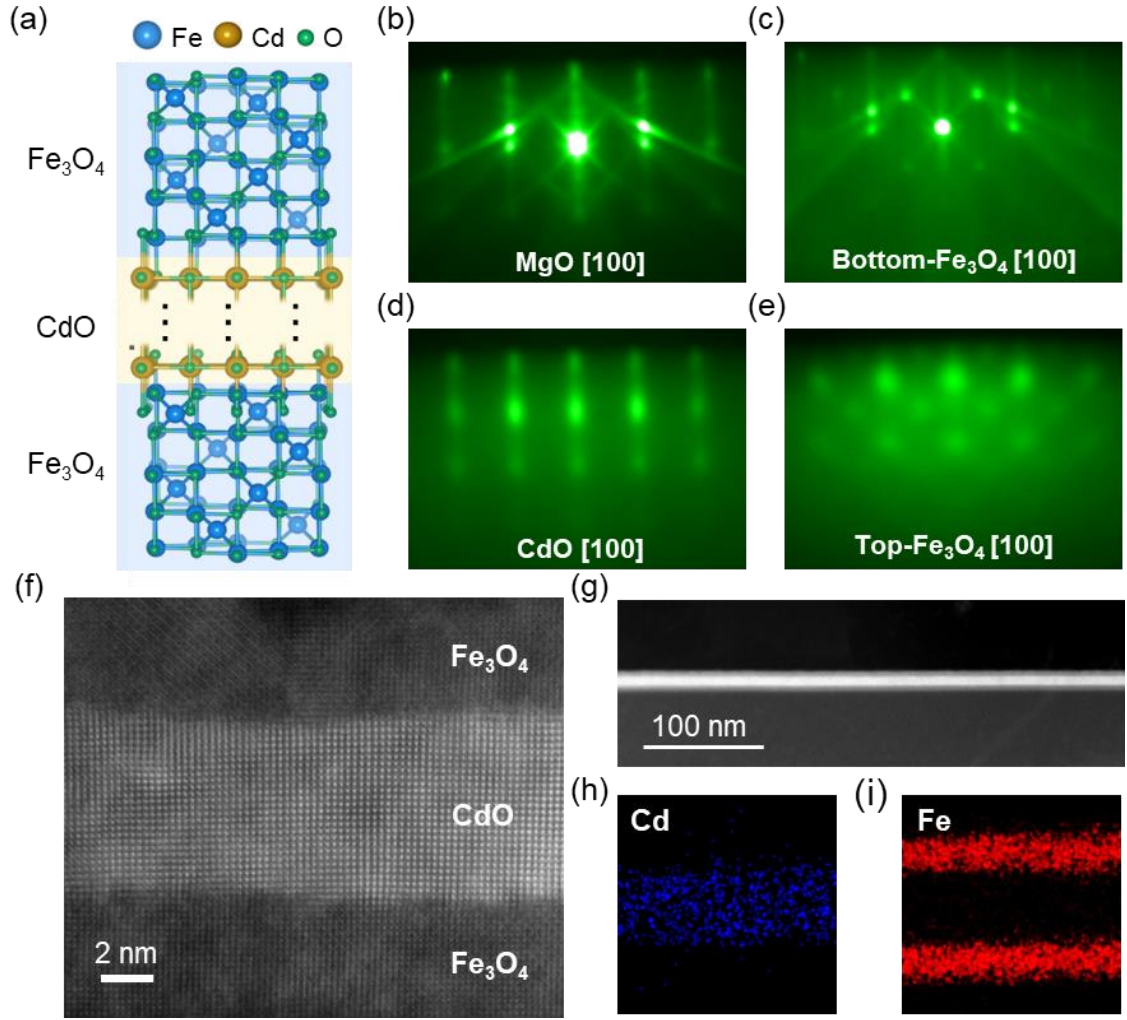

Fig. 1. (a) Schematic of crystalline structures of Fe$_3$O$_4$/CdO/Fe$_3$O$_4$ trilayer superlattices. (b-e) RHEED patterns of (001)-oriented MgO substrate, bottom 4-UC Fe$_3$O$_4$, 15-UC CdO, and top 4-UC Fe$_3$O$_4$ films viewed along the MgO crystal's [100] direction. (f-i) High-angle annular dark field scanning transmission electron microscopy images and EDS of the Fe$_3$O$_4$/CdO (15-UC)/Fe$_3$O$_4$ heterostructures.

**Figure 2**

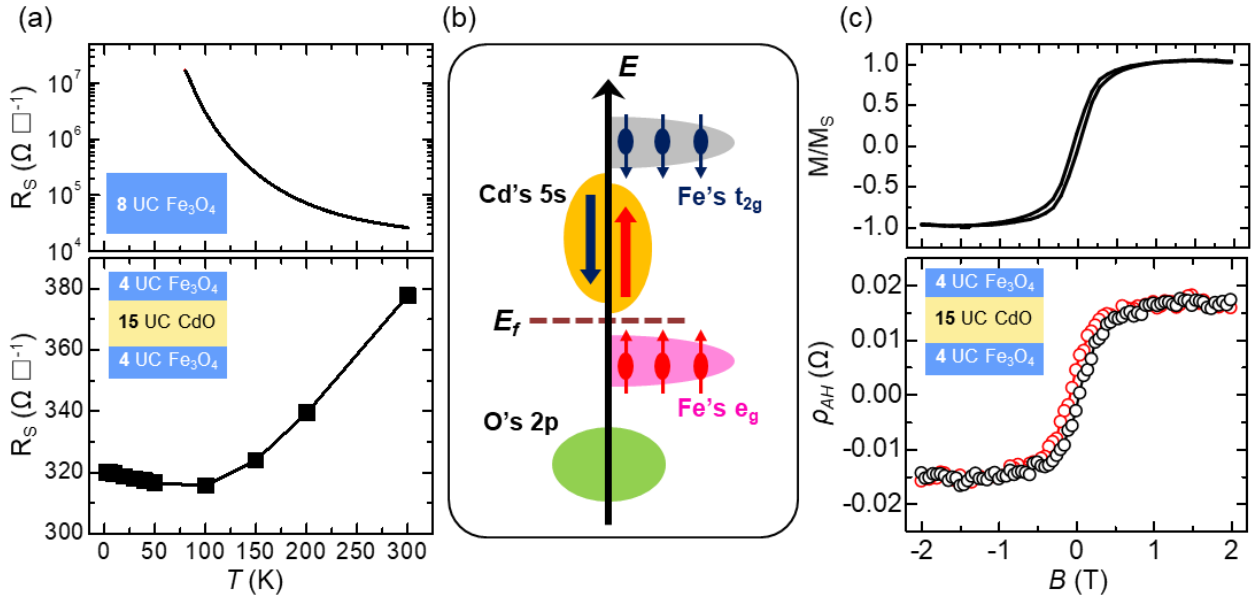

Fig. 2. Anomalous Hall measurements of the $Fe_3O_4/CdO/Fe_3O_4$ trilayer heterostructures. (a) Temperature dependence of sheet resistivity of the 8-UC $Fe_3O_4$ control sample and the $Fe_3O_4$/CdO (15-UC)/$Fe_3O_4$ trilayer sample. (b) Schematic of the magnetic exchange interaction between Cd's *5s* and Fe's *3d* electrons at the CdO and $Fe_3O_4$ interface. (c) Normalized magnetization and the anomalous Hall resistance as a function of the perpendicular magnetic field measured on the $Fe_3O_4$/CdO (15-UC)/$Fe_3O_4$ sample at *T*=10 K.

**Figure 3**

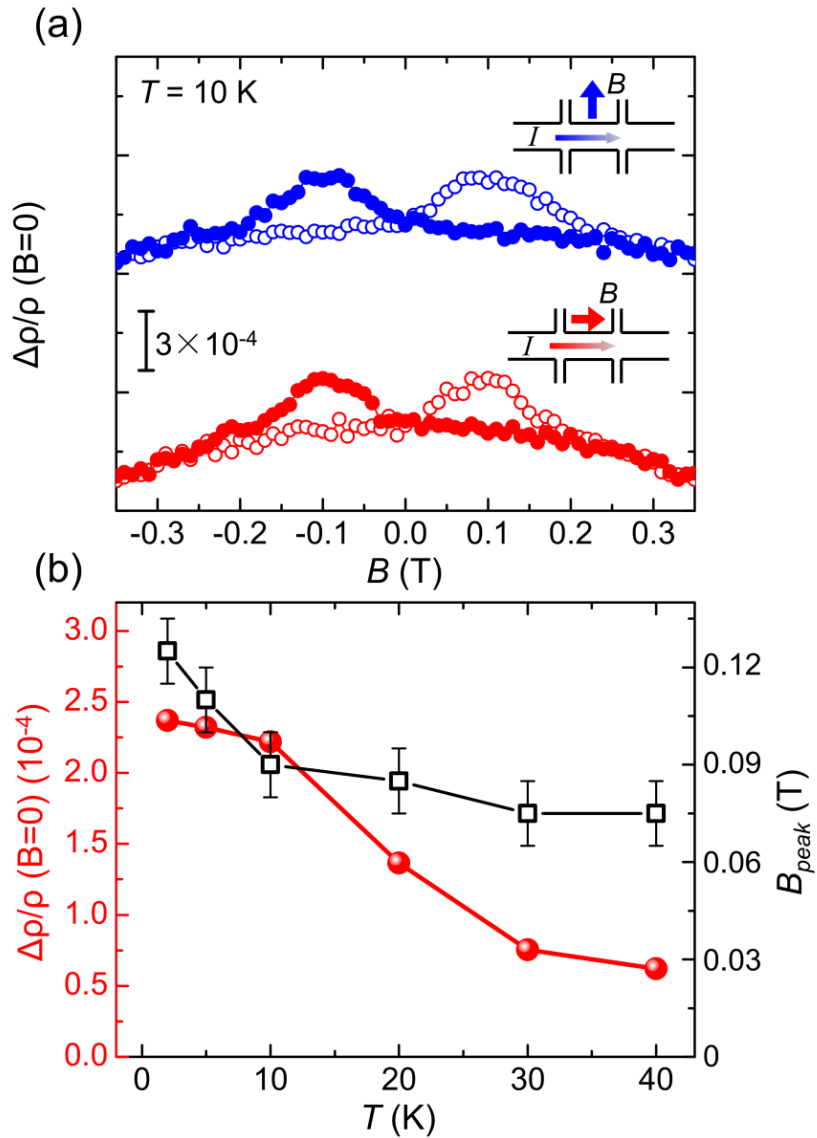

Fig. 3. Planar magnetoresistance measurements of the $Fe_3O_4$/CdO/$Fe_3O_4$ trilayer heterostructures. (a) Planar MR as a function of in-plane magnetic field measured on the $Fe_3O_4$/CdO (15-UC)/$Fe_3O_4$ sample at $T$=10 K. These two curves are shifted for clarity. Inset: Measurement geometries. (b) MR ratio and in-plane magnetic coercive field as a function of temperature.

**Figure 4**

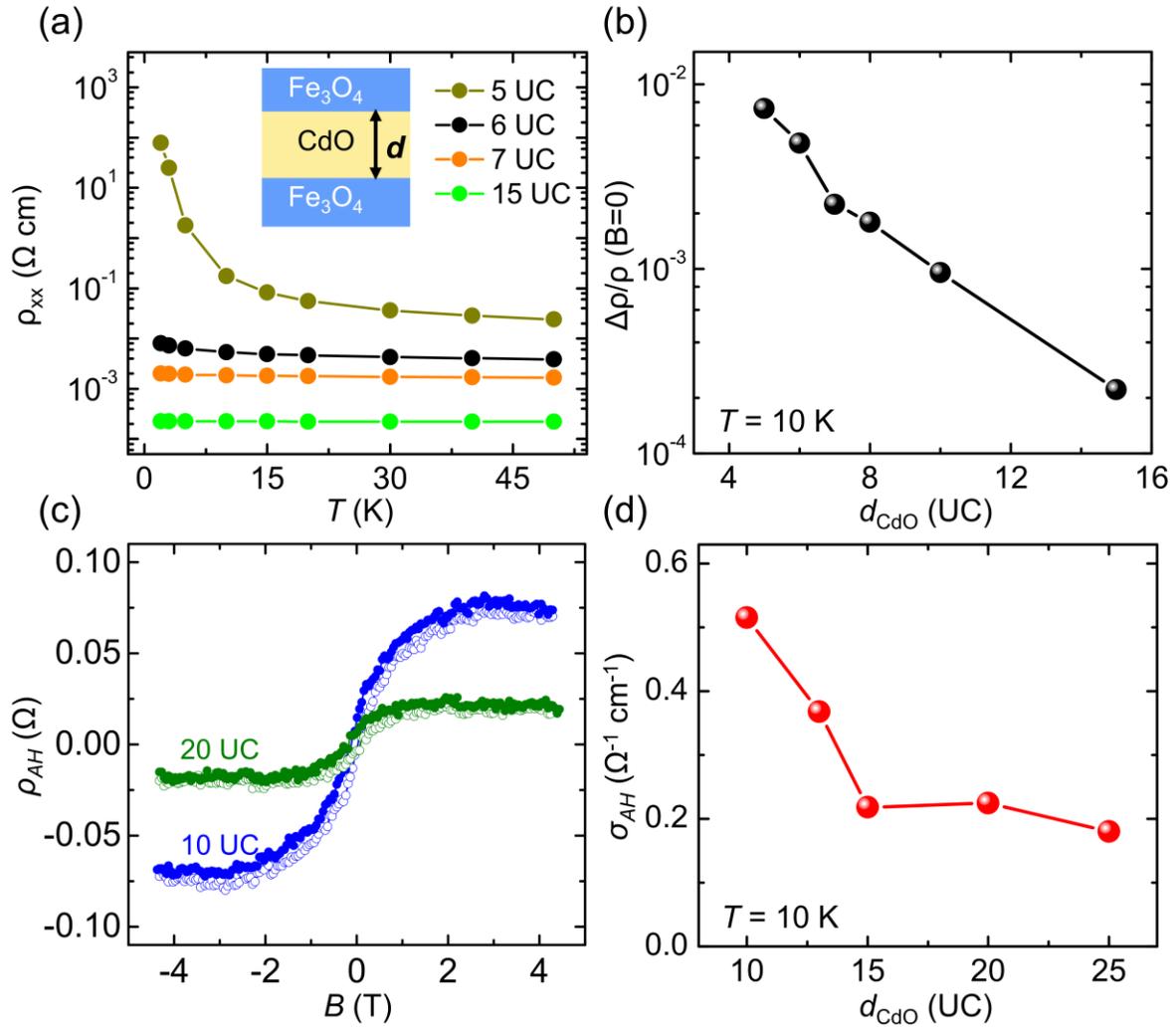

Fig. 4. CdO thickness dependence of the magnetic properties of the $Fe_3O_4$/CdO ($d$-UC)/$Fe_3O_4$ trilayer heterostructures. (a) Channel resistivity versus temperature for several typical $Fe_3O_4$/CdO/$Fe_3O_4$ trilayer samples. (b) MR ratio as a function of CdO thickness at $T$=10 K. (c) Anomalous Hall resistance as a function of perpendicular magnetic field for $Fe_3O_4$/CdO (10 and 20-UC)/$Fe_3O_4$ trilayer heterostructures. (d) CdO thickness dependence of anomalous Hall conductivity at $T$=10 K.

**Supplementary Materials for**

# Interface ferromagnetism and anomalous Hall effect in CdO/ferromagnetic insulator heterostructures

Yang Ma[1,2,†], Yu Yun[1,2,†], Yuehui Li[1,3], Wenyu Xing[1,2], Yunyan Yao[1,2], Ranran Cai[1,2], Yangyang Chen[1,2], Yuan Ji[1,2], Peng Gao[1,2,3], Xin-Cheng Xie[1,2,4,5], and Wei Han[1,2*]

[1]International Center for Quantum Materials, School of Physics, Peking University, Beijing 100871, P. R. China

[2]Collaborative Innovation Center of Quantum Matter, Beijing 100871, P. R. China

[3]Electron Microscopy Laboratory, School of Physics, Peking University, Beijing, 100871, P. R. China

[4]CAS Center for Excellence in Topological Quantum Computation, University of Chinese Academy of Sciences, Beijing 100190, P. R. China

[5]Beijing Academy of Quantum Information Sciences, Beijing 100193, P. R. China

[†]These authors contributed equally to the work

[*]Correspondence to: weihan@pku.edu.cn

# Figure S1

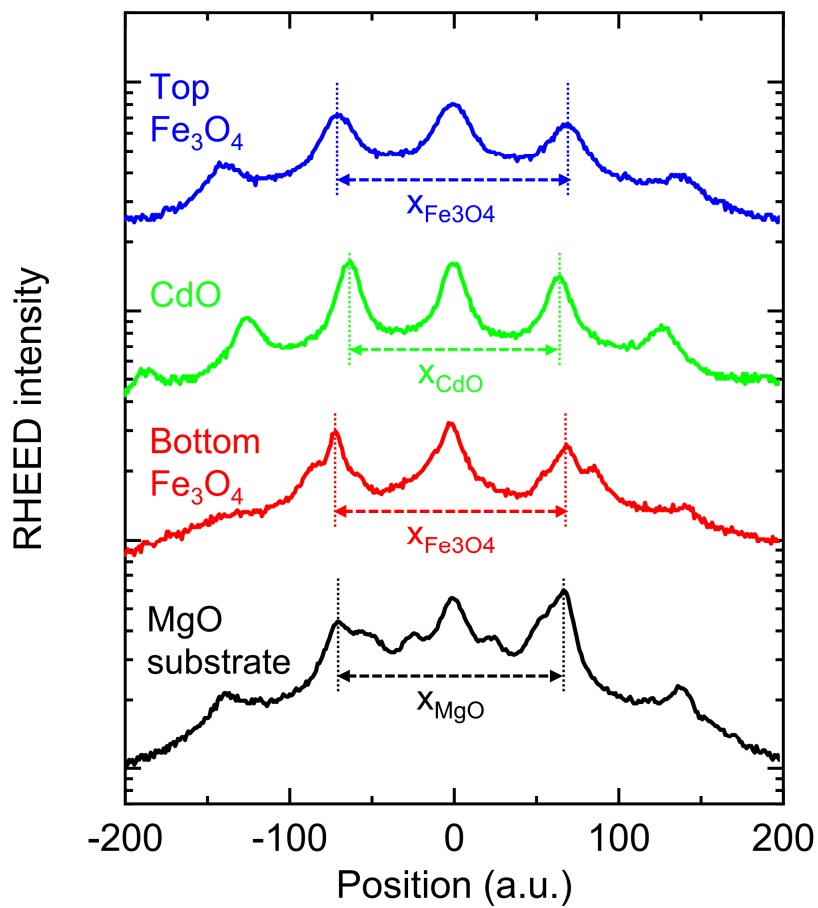

Fig. S1. Line cuts of the RHEED images for the $Fe_3O_4$/CdO (15 UC)/$Fe_3O_4$ heterostructures. Line cuts of RHEED intensity of MgO substrate (black), Bottom $Fe_3O_4$ (red), CdO (green) and Top $Fe_3O_4$ (blue) layers, respectively. The spacing between the major RHEED intensity peaks for each layer are labeled as $X_{MgO}$, $X_{Fe3O4}$ and $X_{CdO}$.

# Table S1

| | Spacing (X) | Bulk lattice constant (nm) | Film lattice constant (nm) | Strain |
|---|---|---|---|---|
| Top $Fe_3O_4$ | 141 | 0.8380 | 0.8185 | 2.33% (Compressive) |
| CdO | 128 | 0.4695 | 0.4508 | 3.98% (Compressive) |
| Bottom $Fe_3O_4$ | 141 | 0.8380 | 0.8185 | 2.33% (Compressive) |
| MgO substrate | 137 | 0.4212 | | |

Table S1. Strain analysis of the $Fe_3O_4$ and CdO layers of the $Fe_3O_4$/CdO (15 UC)/$Fe_3O_4$ heterostructures. The estimation of the strains is obtained based on the line cuts of RHEED intensities (Fig. S1) and using the lattice constant of MgO substrate (a = 0.4212 nm) as a reference.

# Figure S2

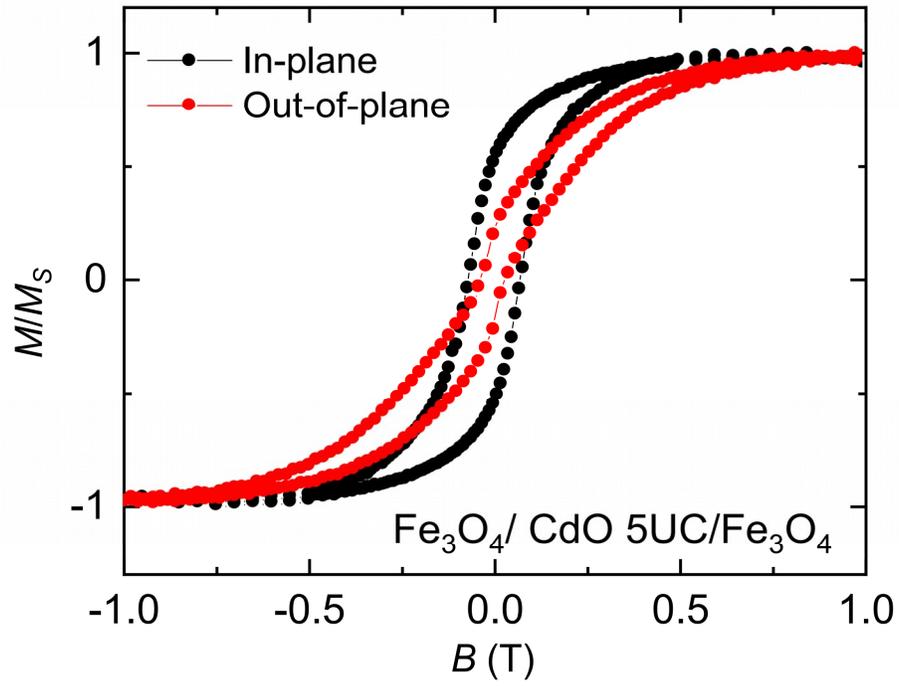

Fig. S2. Typical magnetization measurement of the Fe$_3$O$_4$/CdO/Fe$_3$O$_4$ trilayer heterostructures. The magnetization easy axis is in-plane.

**Figure S3**

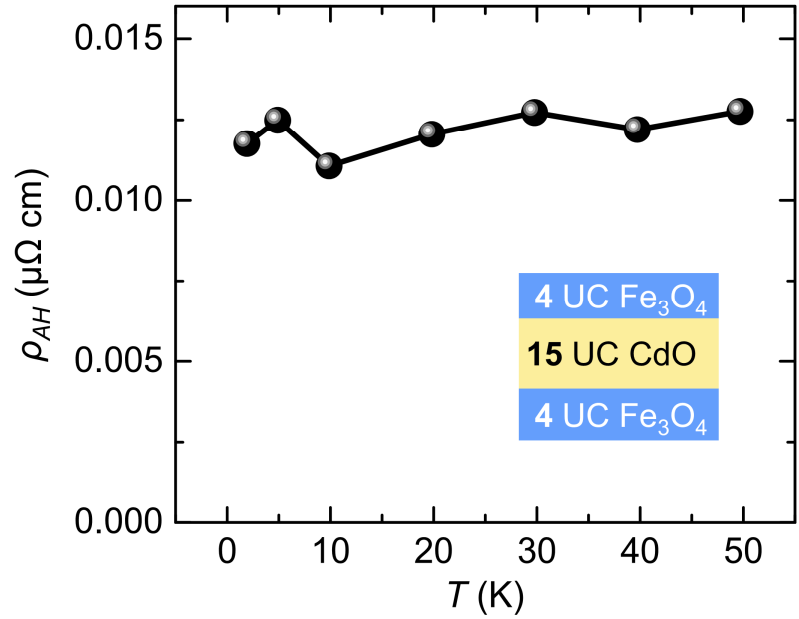

Fig. S3. Temperature dependence of anomalous Hall resistivity measured on the $Fe_3O_4$/CdO (15 UC)/$Fe_3O_4$ trilayer heterostructures.

**Figure S4**

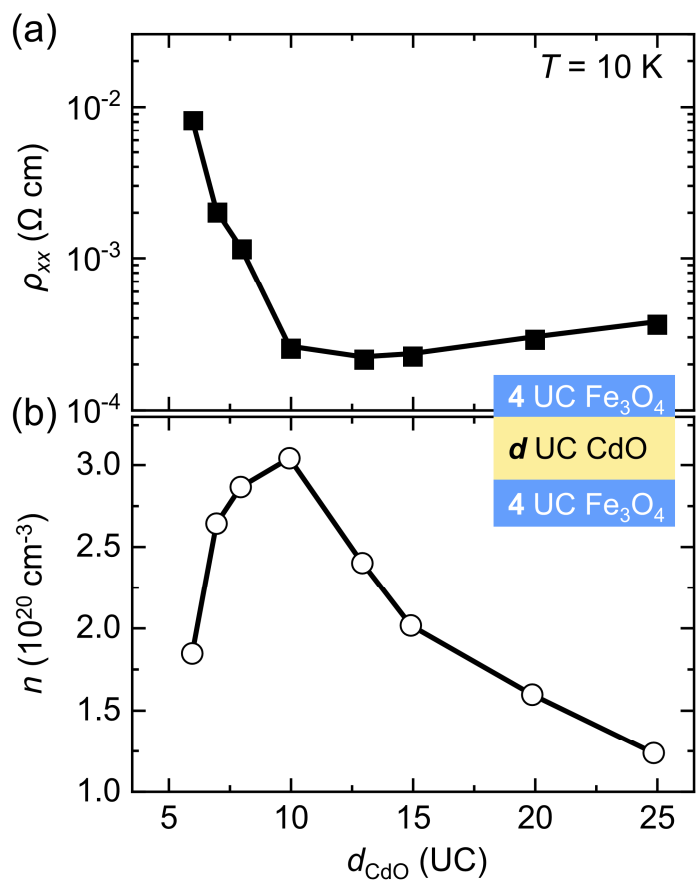

Fig. S4. CdO thickness dependence of the channel resistivity (a) and carrier density (b) for the Fe$_3$O$_4$/CdO ($d$ UC)/Fe$_3$O$_4$ heterostructures. These results are obtained at $T = 10$ K.

**Figure S5**

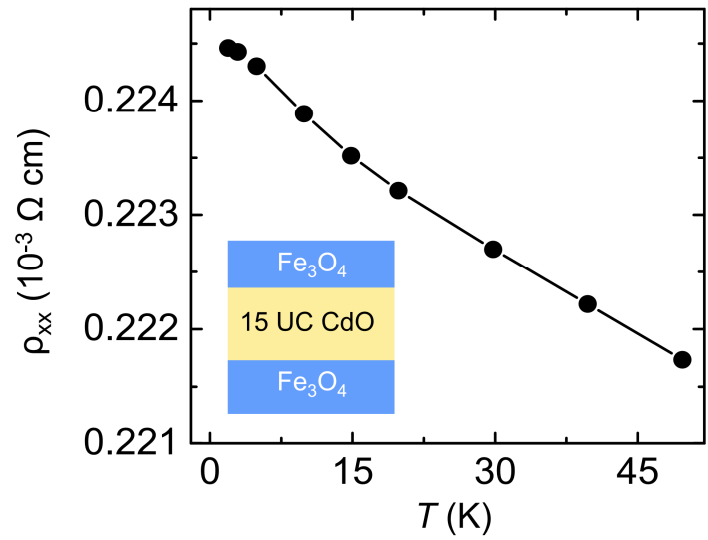

Fig. S5. The temperature dependence of the channel resistivity versus temperature for the Fe$_3$O$_4$/CdO (15 UC)/Fe$_3$O$_4$ trilayer heterostructures. A semiconducting behavior is observed at low temperatures.

**Figure S6**

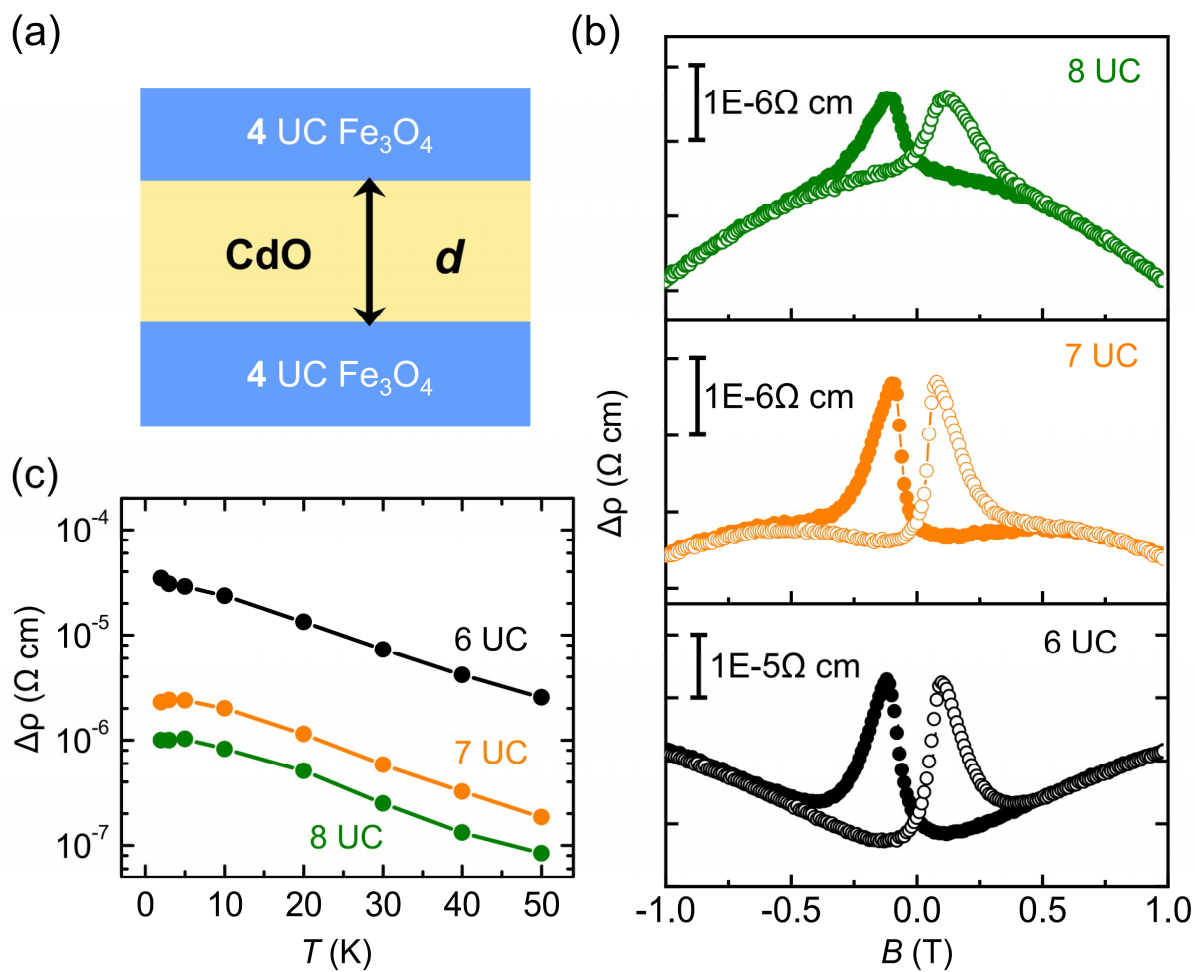

Fig. S6. CdO thickness dependence of the planar MR of the $Fe_3O_4/CdO/Fe_3O_4$ trilayer heterostructures. (a) Schematic of $Fe_3O_4$/CdO ($d$ UC)/$Fe_3O_4$ heterostructures. (b) Planar MR curves as a function of in-plane magnetic field measured on $Fe_3O_4$/CdO (6, 7, and 8 UC)/$Fe_3O_4$ samples. (c) MR ratio ($\Delta\rho/\rho_{B=0}$) as a function of temperature for $Fe_3O_4$/CdO (6, 7, and 8 UC)/$Fe_3O_4$ samples.

**Figure S7**

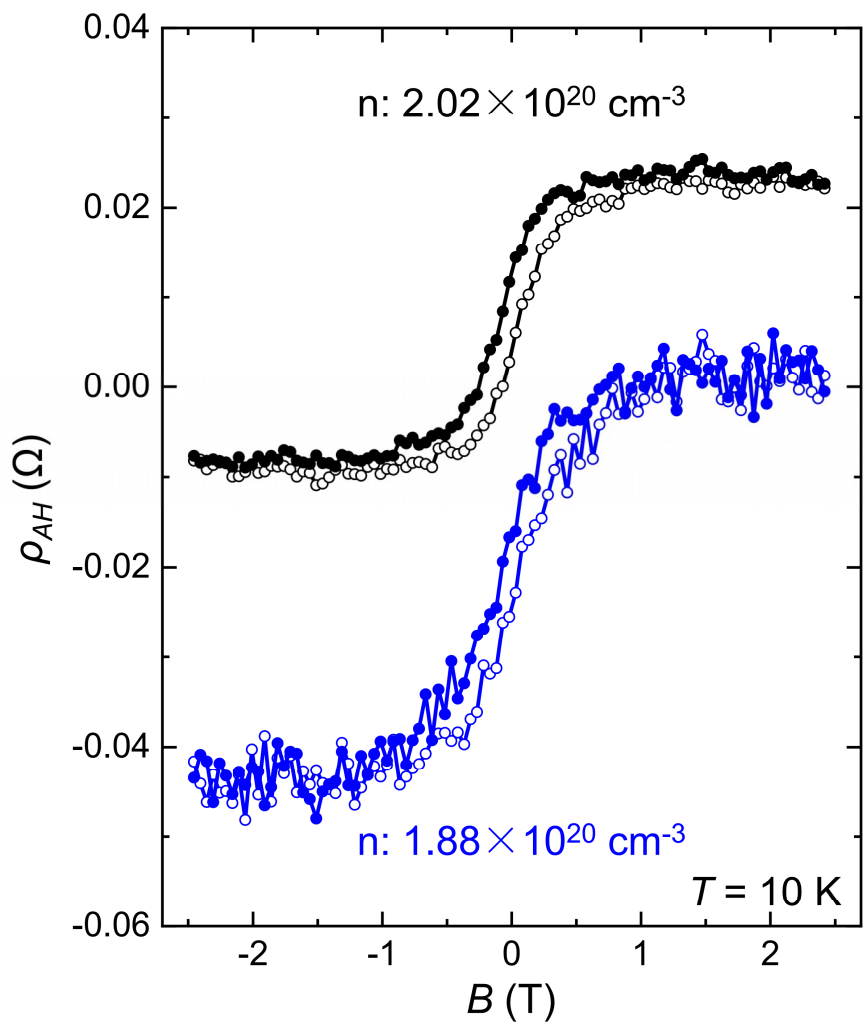

Fig. S7. AHE results of $Fe_3O_4$/CdO (15 UC)/$Fe_3O_4$ heterostructures after decreasing the Cd growth rate and post-annealing in oxygen. A slight larger anomalous Hall resistivity (blue) is observed on $Fe_3O_4$/CdO (15 UC)/$Fe_3O_4$ heterostructures compared to the previous results (black) showed in the main paper (Fig. 2(b)). The carrier density is only slightly reduced (~ 7%).